\documentclass[doublecol]{epl2}
\usepackage[english]{babel}
\usepackage{times}
\usepackage{graphicx}
\usepackage{amsfonts}
\usepackage{psfrag}
\usepackage{verbatim}
\usepackage{color}
\usepackage{url}
\graphicspath{{FIGS/}}

\begin{document}

%opening
\title{Phase separation and critical percolation in bidimensional spin-exchange  models}
\shorttitle{Phase separation and critical percolation}

\author{Alessandro Tartaglia, Leticia F. Cugliandolo and Marco Picco} 

\institute{                    
 Universit\'e Pierre et Marie Curie
  - Paris 6, Laboratoire de Physique Th\'eorique et Hautes Energies,
  4, Place Jussieu, Tour 13, 5\`eme \'etage, 75252 Paris Cedex 05,  France
}

%------------------ABSTRACT----------------------------

\abstract{
Binary mixtures prepared in an homogeneous phase and quenched  into a two-phase 
region phase-separate via a coarsening process whereby domains of the two phases grow in time. With a numerical 
study of a spin-exchange model we show 
that this dynamics first take a system with equal density of the two species to a critical percolation state. 
We prove this claim  and we determine the time-dependence of the growing length associated to this process
with the scaling analysis of the statistical and morphological properties of the clusters of the two phases.
}

%%%%%%% \keywords{Suggested keywords}%Use showkeys class option if keyword

\maketitle

%%%%%\setlength{\textfloatsep}{10pt} 
%%%%%\setlength{\intextsep}{10pt}

%%%%%%%%%%%%%%%%%%%%%%%%%%%%%%%%%%%%%%%%%%%%%
%\section{Introduction}
%%%%%%%%%%%%%%%%%%%%%%%%%%%%%%%%%%%%%%%%%%%%%

Phase separation is the process whereby a binary mixture
of components A and B, initially in a homogeneous phase,
demix. This  process leads to the coexistence of two phases: one rich in A
and the other in B~\cite{Bray94,Bray03,Puri09-article,GonnellaYeomans09,Tanaka12,CorberiPoliti}. 
The system, initially in an unstable spatially
uniform state, progressively coarsens to approach
its thermodynamically stable phase-separated state. Such phenomena arise in binary alloys, fluid mixtures, and
polymer blends. Recently, the dynamics of phase separation have seen a revival of interest  
in the context of experimental~\cite{De-etal14,Tojo-etal} and numerical~\cite{Hoffman-etal14,KudoKawaguchi,Takeuchi15,Takeuchi16} 
studies of binary mixtures of Bose gases.

The  {\it late} time dynamics are well understood. 
In the absence of driving forces, a 
dynamic scaling regime with statistically 
self-similar domain morphology sets in. This regime is well-described by an extension of the 
Lifshitz-Slyozov-Wagner (LSW) theory~\cite{LifshitzSlyozov59,Wagner61}, in which the typical 
domain radius grows as~\cite{Hu86}
\begin{equation}
\ell_d(t) \simeq t^{1/z_d}
\qquad
\mbox{with}
\qquad
 z_d=3
 \end{equation}
(whereas for scalar non-conserved order parameter dynamics 
the growing length is also given by a power law but the exponent is $z_d=2$). Numerical results in 
favour of this law were published in~\cite{Hu86,Amar88,Rogers88} for spin-exchange models 
although the 
growth-law can be more complex in particle or polymer phase separating systems,
see {\it e.g.}~\cite{Reith12} and references therein. The pre-asymptotic dynamics 
leading to this regime have not been discussed in detail in the literature.

It was noticed in~\cite{SiSaArBrCu09} that the low-temperature evolution 
of a bidimensional 50:50 binary mixture after a quench from infinite temperature shares many points in common with the 
one generated by Glauber single spin-flip stochastic dynamics satisfying detailed balance~\cite{ArBrCuSi07,SiArBrCu07}. On the one hand, an {\it early} approach to 
critical percolation was noticed, although the time needed to reach this 
state was not studied in detail. On the other hand, a separation of length-scales in the statistics and 
morphology of finite size cluster areas and domain wall lengths was observed. 
Linear or planar objects that are smaller than the typical ones, $\ell_d(t)$ or $\ell_d^2(t)$, 
satisfy dynamic scaling with respect to $\ell_d(t)$, while larger objects were found to be very 
close to the ones of critical percolation.

In this Letter we characterise the early stages of the dynamical process. 
More precisely, we analyse the way in which the system approaches a state with a 
stable pattern of critical percolating domains. We monitor a number of observables (to be defined in the main part of the text)
and we explain how their behaviour constitutes evidence for this claim.
We prove the approach to critical percolation for balanced mixtures whereas 
different behaviour is found for asymmetric ones~\cite{Takeuchi15}. 

\begin{figure}[h]
\vspace{0.5cm}
\begin{center}
        \includegraphics[scale=0.39]{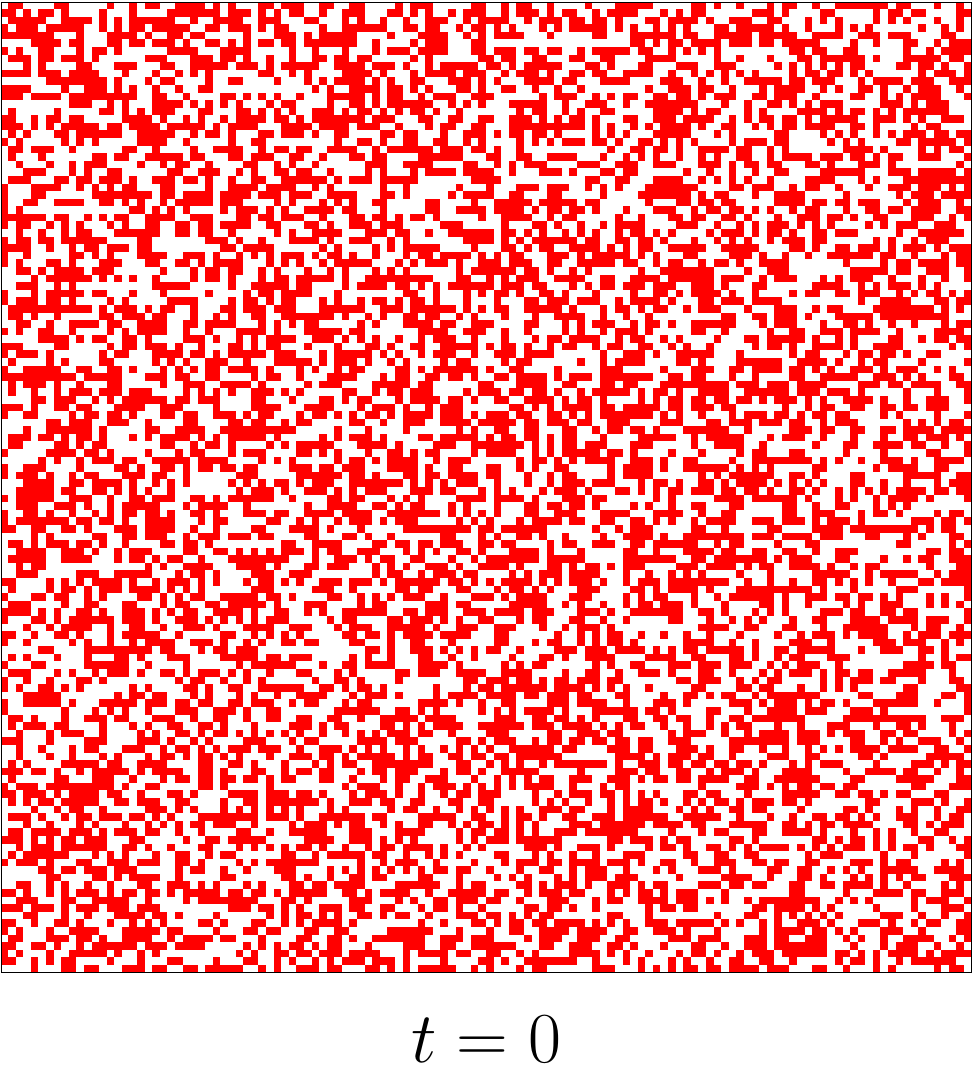}\quad%
	\includegraphics[scale=0.39]{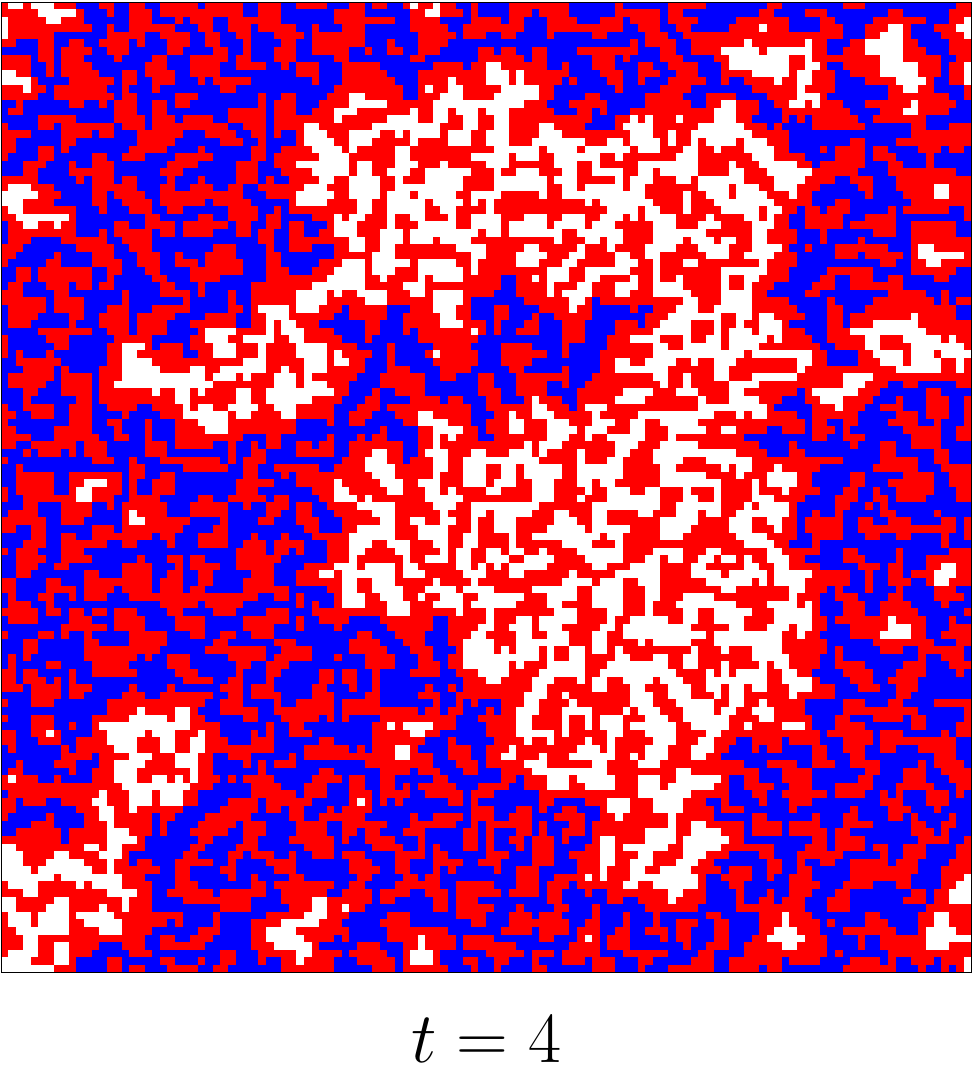}\quad%
        \includegraphics[scale=0.39]{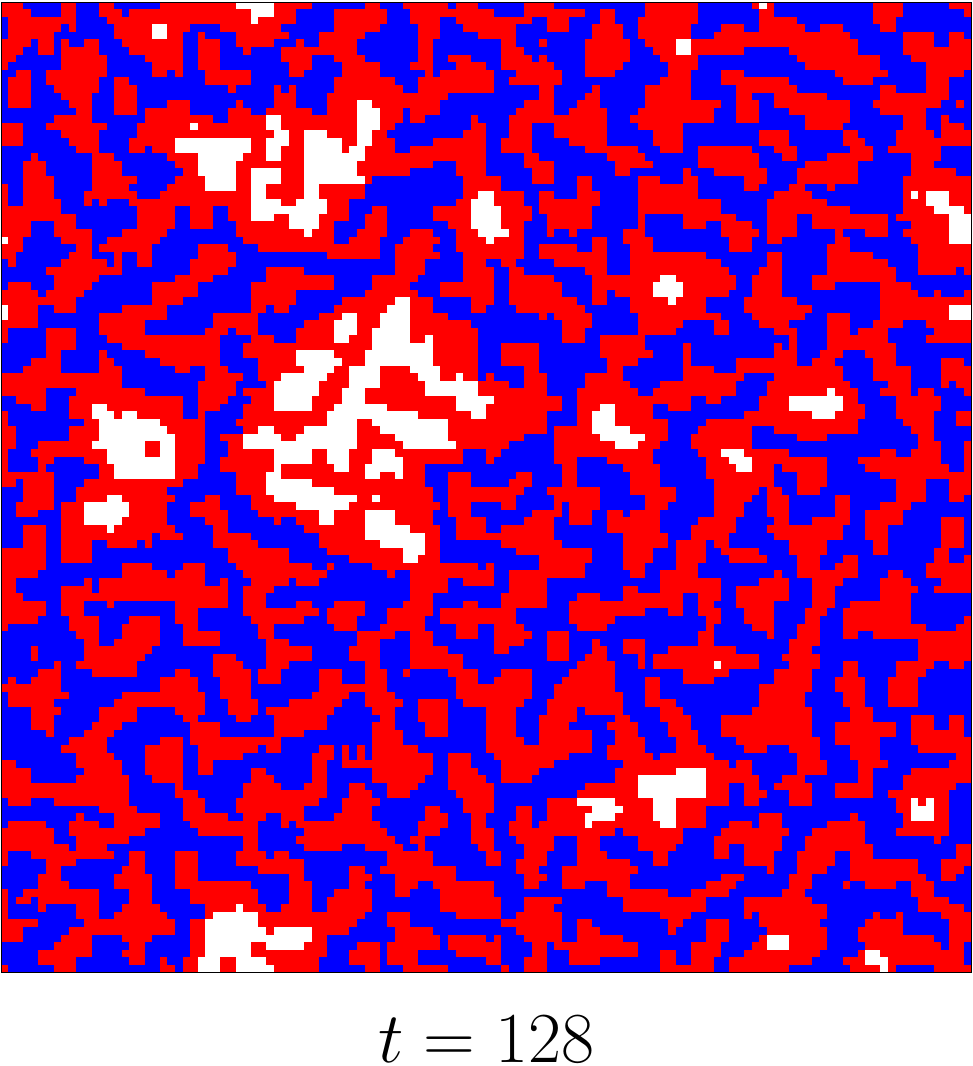}\quad%
	\includegraphics[scale=0.39]{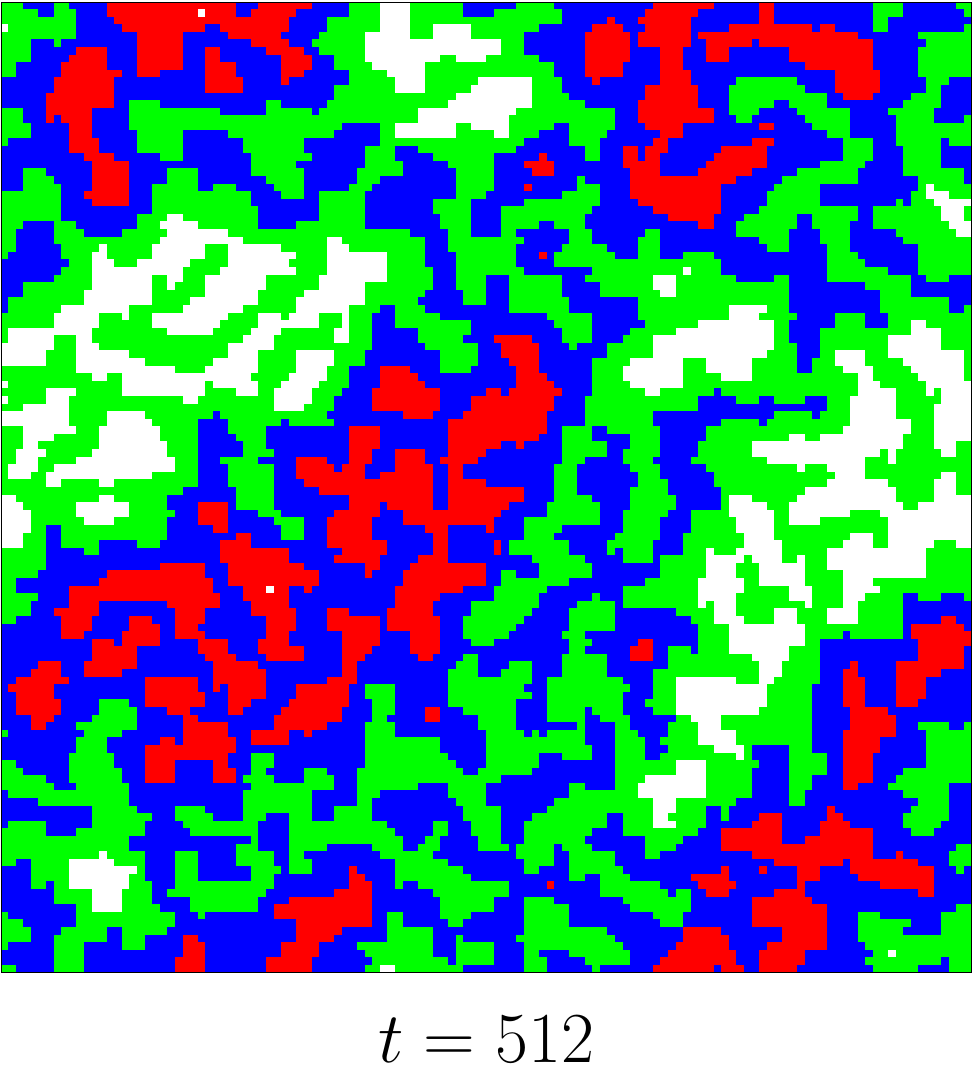}\quad%

\end{center}
\caption{\small Instantaneous spin configurations on
an $L=128$ square lattice with periodic boundary conditions.
Red sites and white sites represent $+1$ and $-1$ spins, respectively. 
Clusters that wrap around the system are highlighted in 
green for spin $+1$ wrapping clusters, and blue for $-1$ wrapping clusters. }
\label{KaSnap}
\end{figure}

One of the quantitative goals of this work is to measure the 
growing length,  $\ell_p(t)$,  that characterises the approach to critical percolation for phase separating systems
on the lattice. Accordingly, we define the time needed to reach critical percolation, 
$t_p$, from $\ell_p(t_p) = L$ with $L$ the linear size of the sample.
For Glauber dynamics the approach to the percolating state 
is characterised by 
$\ell_p(t) \simeq t^{1/z_p}$, with $z_p < z_d $, and the numerical analysis 
suggested~\cite{BlCoCuPi14}
\begin{equation}
\ell_p(t) \simeq \ell^n_d(t)
\label{eq:lp-ld}
\end{equation} 
with $n$ the coordination of the lattice. The results that we will present here
indicate that such an algebraic relation is also satisfied
for the local spin exchange (Kawasaki) dynamics of binary mixtures with 
 equal concentration of the two species.
For Glauber dynamics, reasoning in terms of $z_p$ and $z_d$ 
 gives very good  results, since $z_d=2$ over a  long time-interval. However, for locally conserved order parameter dynamics
 the power-law is hard to establish  and the representation 
 of $\ell_p$ in terms of $\ell_d$, as in Eq.~(\ref{eq:lp-ld}),  is better. As for the exponent $n$, we argued  that, in Glauber dynamics, 
 it is the coordination of the lattice that we can associate to half the maximal possible energy change induced by a single spin flip~\cite{BlCoCuPi14}. 
For local Kawasaki dynamics neighbouring pairs of spins are updated 
simultaneously. Such a change can induce, at most, an energy change equal to twice the number of nearest-neighbours of 
a neighbouring pair of sites. If we follow the argument for the Glauber case, $n=6$ for the triangular lattice (we built this lattice in 
such a way that there are  8 nearest-neighbours of a pair of sites but two of them are shared by the two spins in the pair), 
$n=6$ for the square lattice, 
and $n=4$ for the honeycomb lattice. Note, however, that the triangular lattice is special as the 50:50 initial 
conditions are right at the critical percolation point. We will discuss this guess in the body of this Letter.
 
\begin{figure}[h]
\vspace{0.5cm}
\begin{center}
        \includegraphics[scale=0.18]{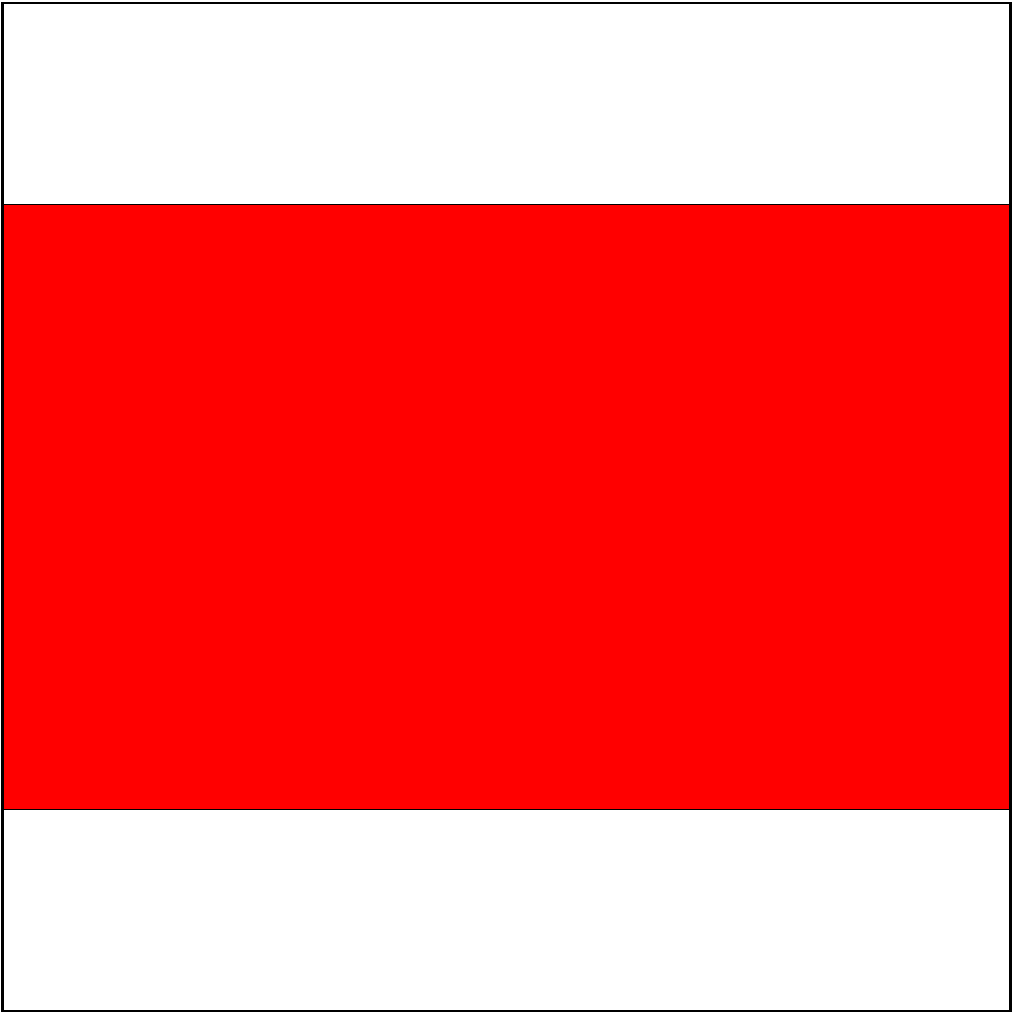}
         \includegraphics[scale=0.18]{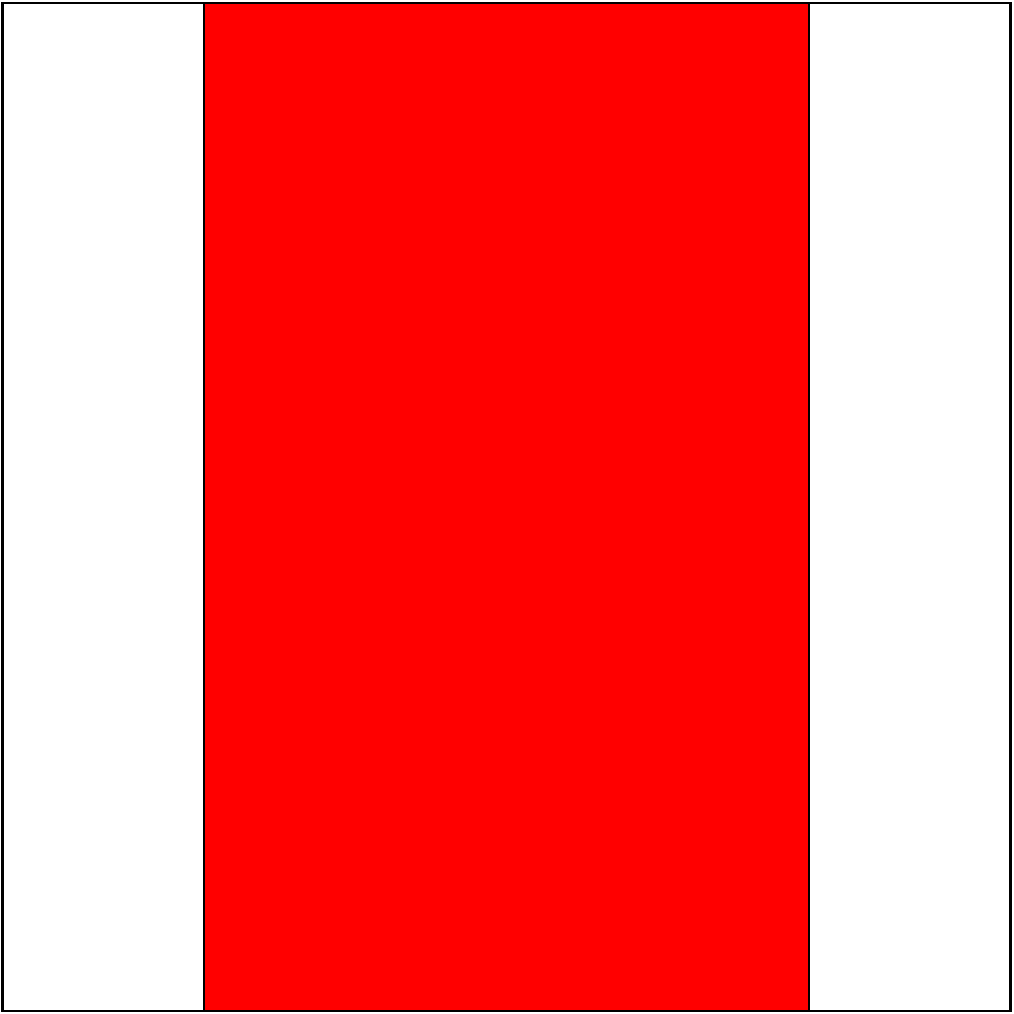}
         \includegraphics[scale=0.18]{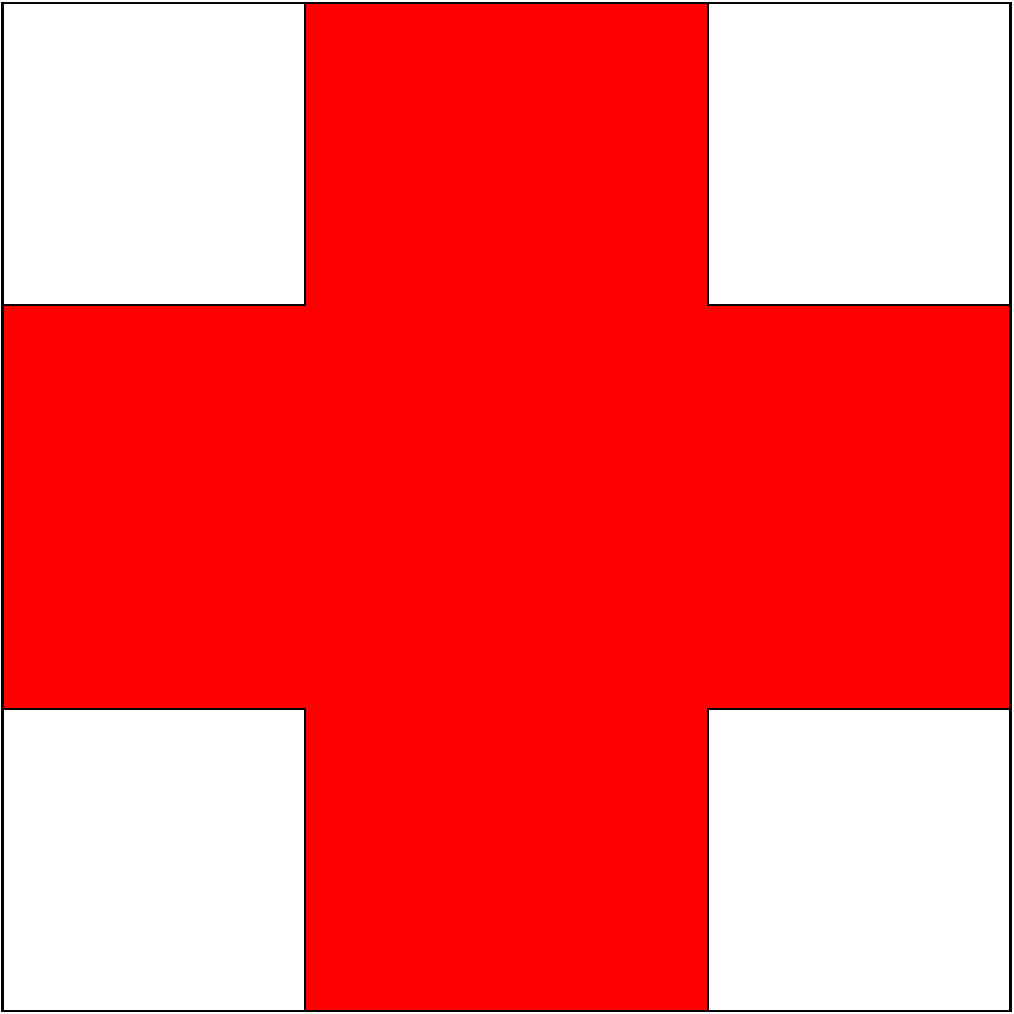}
         \includegraphics[scale=0.18]{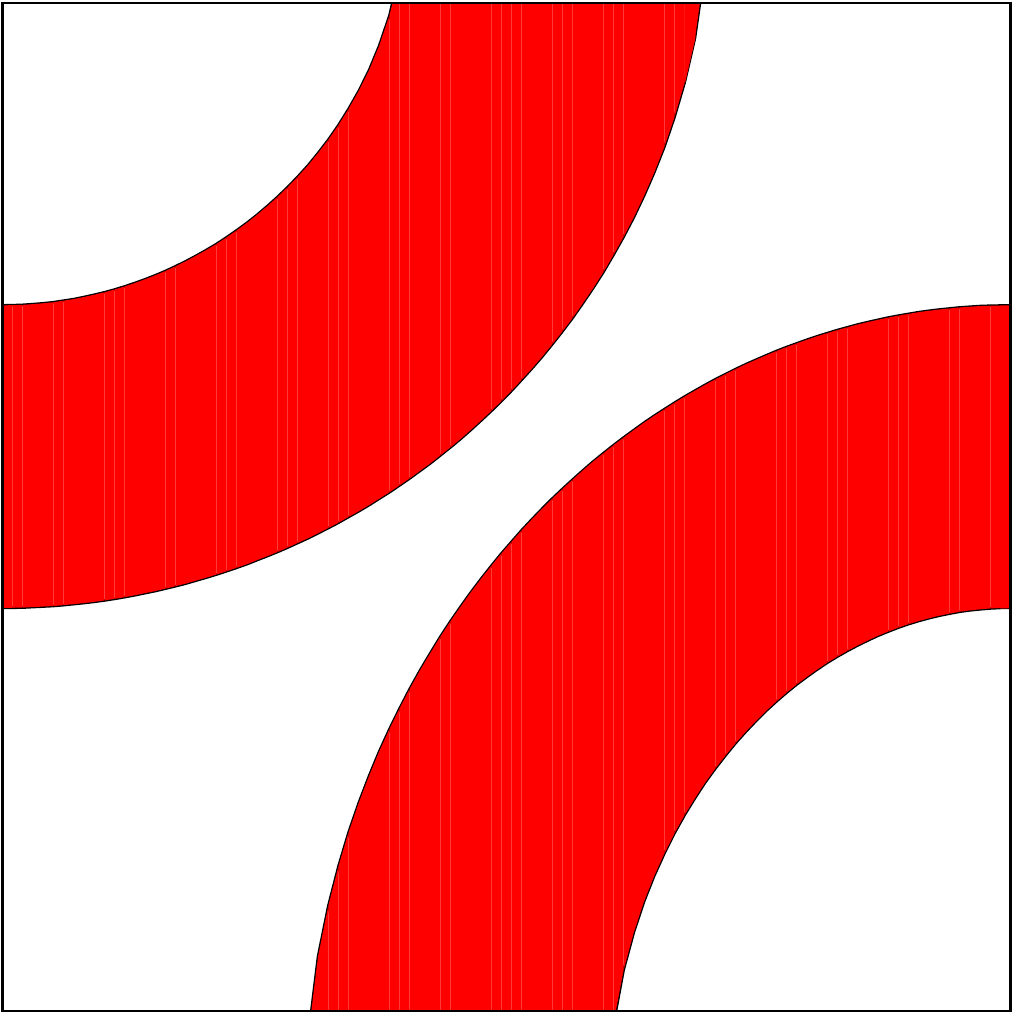}
         \\
         h $\qquad\qquad\quad$ v $\qquad\qquad\quad$ hv $\qquad\qquad\quad$ diag
\end{center}
\caption{\small 
Sketch of configurations with percolating clusters on a square lattice
with periodic boundary conditions (a torus).
}
\label{fig:sketch}
\end{figure}

Concretely, we study a lattice gas model with 
local Kawasaki rules~\cite{Kawasaki66a,Kawasaki66b,Bortz-etal74,Barkema} used to mimic 
phase separation in systems in which hydrodynamic effects can be neglected~\cite{Bray03}. 
We use the spin language, in which 
up and down Ising variables correspond to the presence of the A and B 
species on a given site. The energy function is the familiar ferromagnetic Ising one
\begin{equation}
H = -J \sum_{\langle ij\rangle} s_i s_j
\end{equation}
with the sum running over pairs of nearest neighbour sites on the lattice and $J>0$.
We set $J=1$ so that the critical temperature on a square lattice is 
$T_c = 2 \, \left[ \ln{(1+\sqrt{2})} \right]^{-1}$ in units of $k_{B}$.
The spin exchange (Kawasaki) dynamic rules are defined as follows.
At each time step, a pair of nearest-neighbour sites is chosen at random.
If the spins sitting on these sites are antiparallel they are exchanged 
with the heat-bath Monte Carlo rule.  
If the two sites are occupied by the  same kind of spin their state remains unchanged. 
The control parameters are temperature 
and the relative concentration of the two species.
All data shown are for the Ising model on a square lattice  with periodic boundary conditions
after quenches from infinite temperature to $T_c/4$ and $T_c/2$. Results on other lattice geometries will be mentioned but 
not shown.

\begin{figure}[h]
\begin{center}
        \includegraphics[scale=0.64]{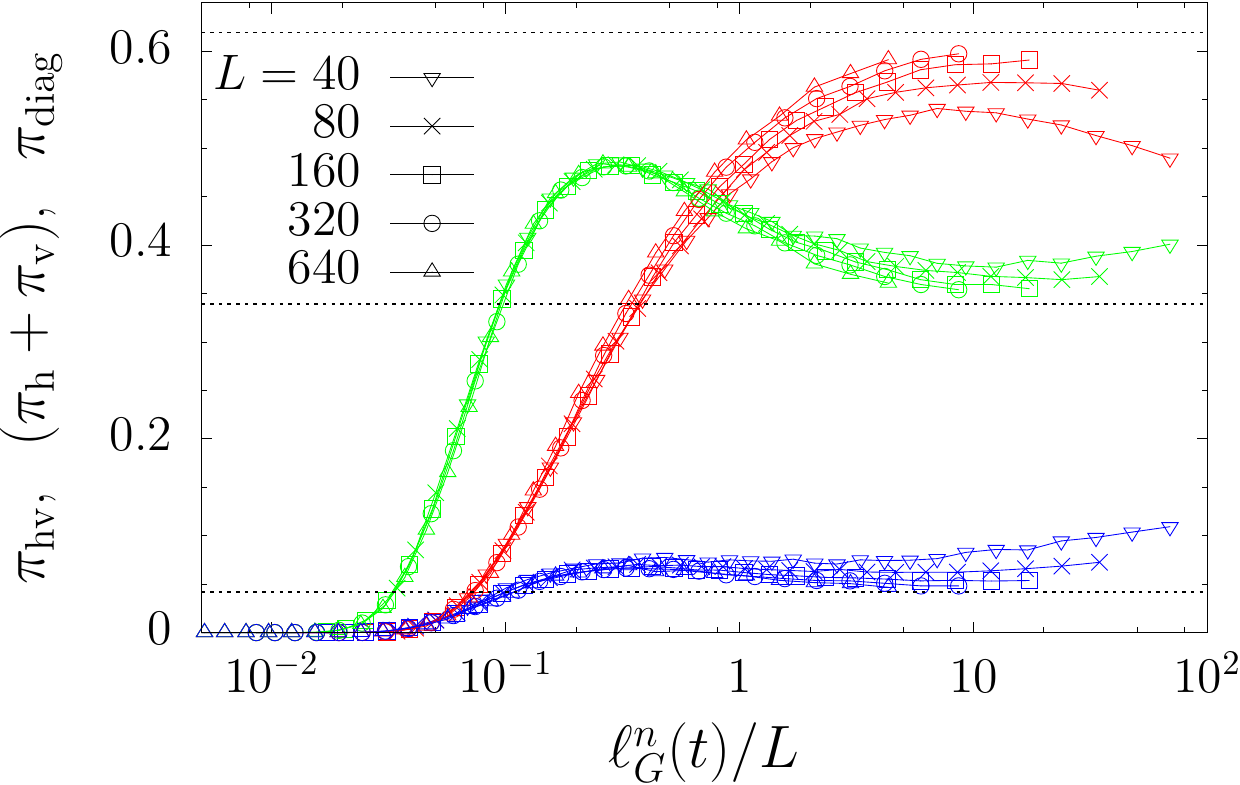}
        \\
        \vspace{0.5cm}
        \includegraphics[scale=0.57]{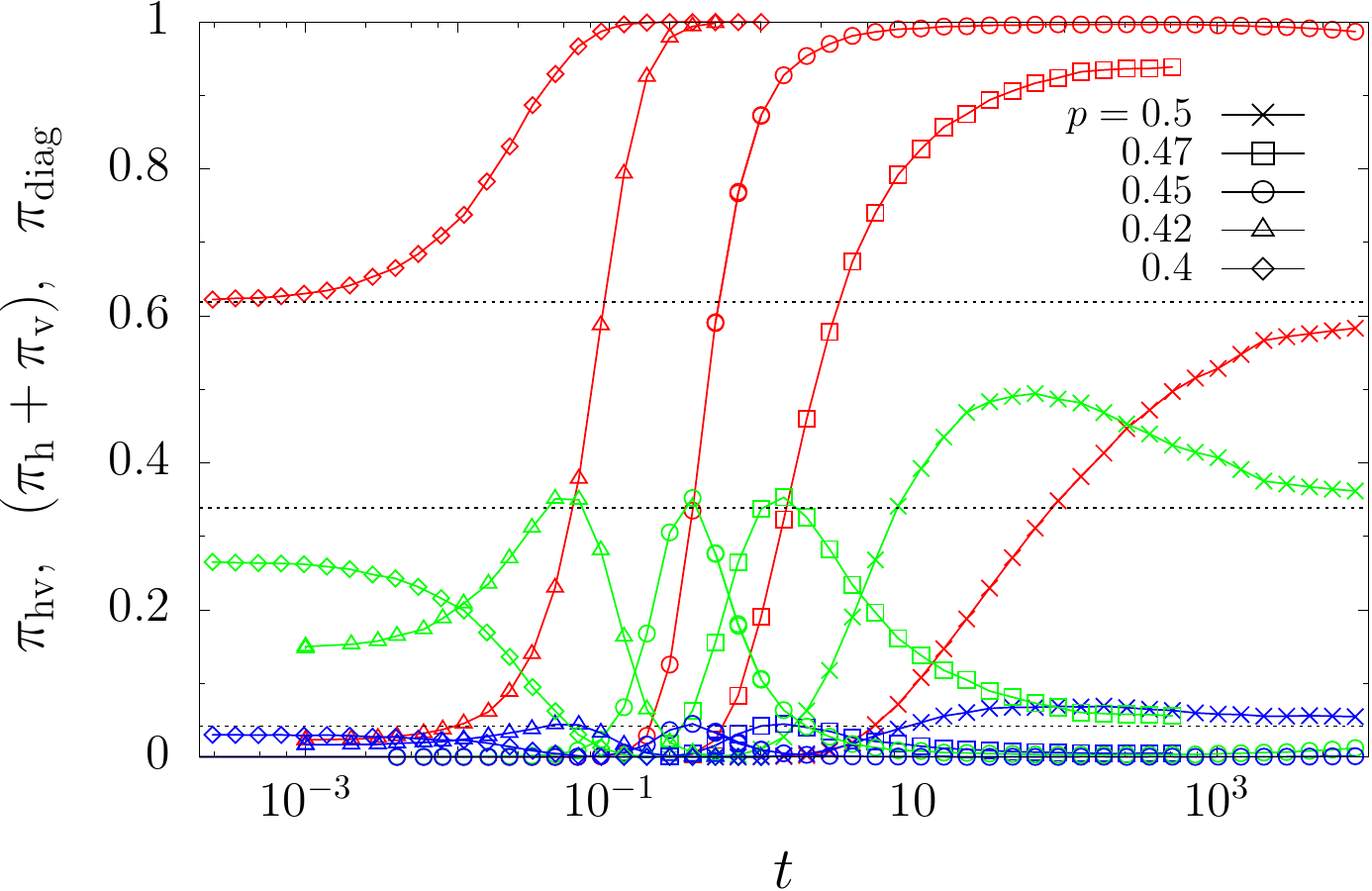}
\end{center}
\caption{\small
Wrapping probabilities  after quenches to $T_c/2$. Upper red curves: clusters wrapping in both principal directions of the lattice ($\pi_{{\rm hv}}$).
Intermediate green curves: clusters wrapping in only one principal direction, either horizontally or vertically ($\pi_{{\rm h}} + \pi_{{\rm v}} $).
Lower blue curves: clusters wrapping in a diagonal direction ($\pi_{{\rm diag}}$).
The horizontal dotted lines are the exact values of the wrapping probabilities for critical percolation.
Upper panel: data for systems with equal concentration of up and down spins
as a function of the scaling variable $ \ell^n_G(t)/L$ with $n=5$ and various sizes given in the key.
Lower panel:  system with linear size $L=160$ and fixed concentration $p = 0.5, \ 0.47, \ 0.45,\ 0.42, \ 0.4$ 
of $+1$ spins. 
}
\label{LKaWrapping_b2}
\end{figure}

In Fig.~\ref{KaSnap} we show the evolution of the 
characteristic domain pattern in a system with linear size $L=128$. 
The concentration of up (red) and down (white) spins is a half. The clusters that percolate in at least one direction 
are highlighted in green and blue,
respectively. Very early, already at $t=4$, a percolating cluster appears, then breaks (not shown) and rebuilds again
($t=128$) until in the late-time snapshot, at $t \geq 512$,  {\it two} large clusters of opposite orientation are interlaced and 
percolate horizontally. This configuration belongs to the first class sketched in Fig.~\ref{fig:sketch}, named `${\small \rm h}$'
for horizontal along the torus.
Note that other runs can lead to configurations of the `v'  and `diag' type with also two percolating clusters or `hv' with only one 
percolating cluster. We will show the probability of reaching each of these in Fig.~\ref{LKaWrapping_b2} and discuss them 
in the text below.

\begin{figure}[h]
\vspace{0.5cm}
\begin{center}
        \includegraphics[scale=0.6]{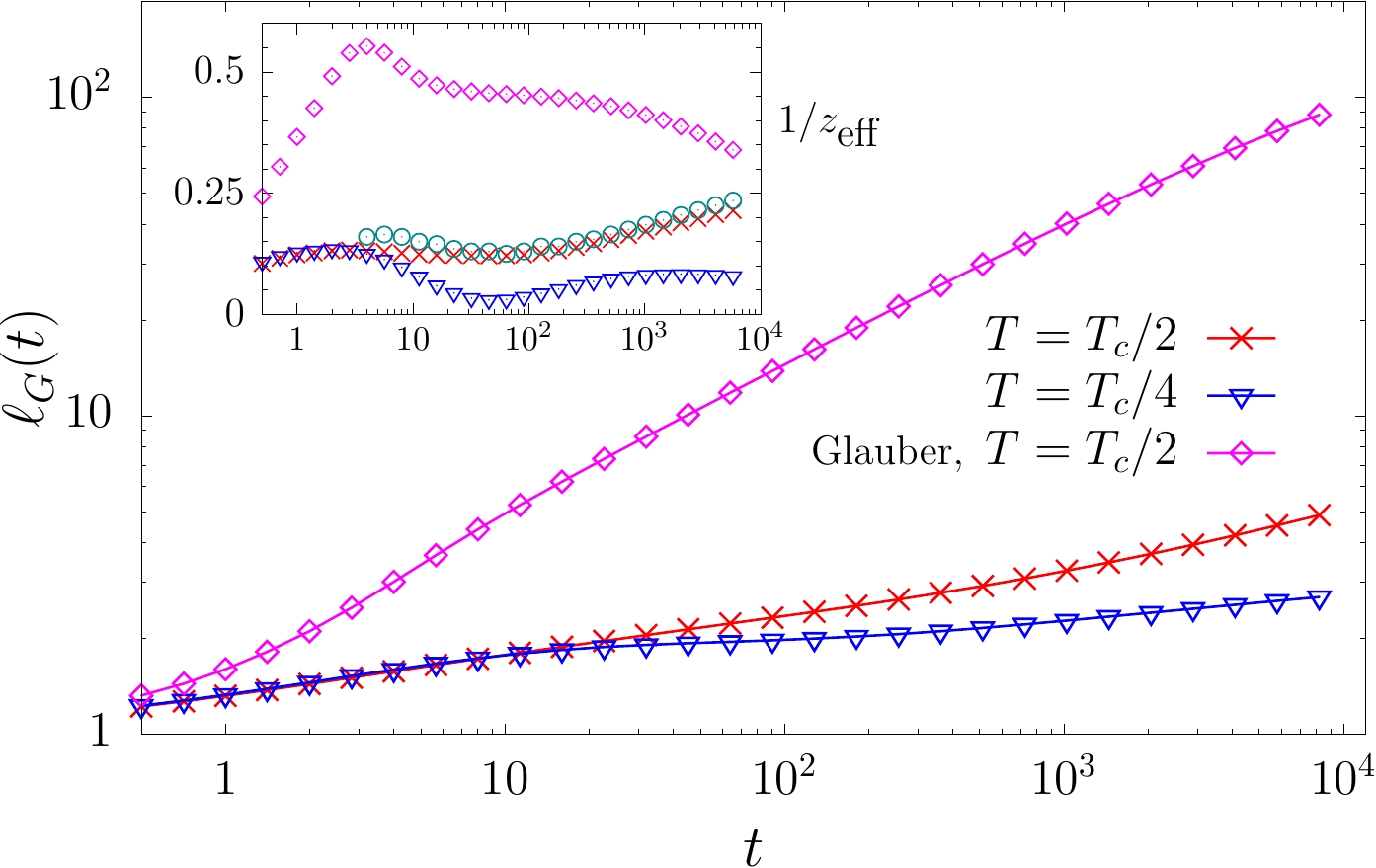}
\end{center}
\caption{\small 
Time-dependent growing length $\ell_G(t)$, defined in Eq.~(\ref{eq:growing-length-excess-energy}),
following a sudden quench 
to $T=T_c/2$ (red crosses) and $T=T_c/4$ (blue down triangles) and, for comparison, 
data for Glauber dynamics at $T_c/2$ (purple diamonds). System with $L=640$
and balanced densities of the two species.
In the inset, the effective growth exponent, $1/z_{\mathrm{eff}}(t)$, computed as the 
logarithmic derivative of $\ell_G(t)$ (same symbols as in the main part of the figure)
and the scaling of the variance of the winding angle, $\langle \theta^2 \rangle$ (open green circles), 
see Eq.~(\ref{ws}) and the main text.}
\label{LKaGL}
\end{figure}

A way to estimate the time-dependence of the dynamic growing length involved in the dynamic scaling is to use
the time-dependent excess energy, $E(t) =  \langle H(t)\rangle$, with respect to  the ground state energy $E_0$ 
\begin{equation}
\ell_{G}(t) = E_{0} /[  E_{0} - E(t)]
\label{eq:growing-length-excess-energy}
\; .
\end{equation}
(e.g.,  $E_{0}= - 2 L^2 \, J$ for the square lattice).
The angular brackets indicate an average over different realisations of the dynamics (initial conditions and/or thermal noise).
This quantity is shown in Fig.~\ref{LKaGL} for Kawasaki and Glauber dynamics at two sub-critical temperatures, 
$T_c/2$ and $T_c/4$. Contrary to what happens for Glauber dynamics, 
nowhere in the time span shown
in the figure a stable algebraic increase of $\ell_G$ established. 
The evolution of the effective exponent in 
$\ell_G(t) \simeq t^{1/z_{\rm eff}(t)}$ is followed in the inset. In this time-window 
$z_{\rm eff}(t)$ goes roughly from $10$ to $4$ for $T_c/2$, and the measurement slowly approaches the 
expected value for the dynamic exponent $z_d =3$~\cite{Hu86} but time 
scales of two orders of magnitude longer are needed to reach convergence~\cite{Amar88,Rogers88}.
For other geometries, such as triangular or honeycomb lattices,
the values of $z_{\rm eff}$ are similar to the ones for the square lattice~\cite{Tartaglia-long}.
The fact that the effective exponent varies so much in time and depends on temperature 
suggests to use the time-dependent growing length itself to analyse the pre-asymptotic regime
with the eventual approach to percolation. 

\begin{figure}[h]
\vspace{0.5cm}
\begin{center}
        \includegraphics[scale=0.6]{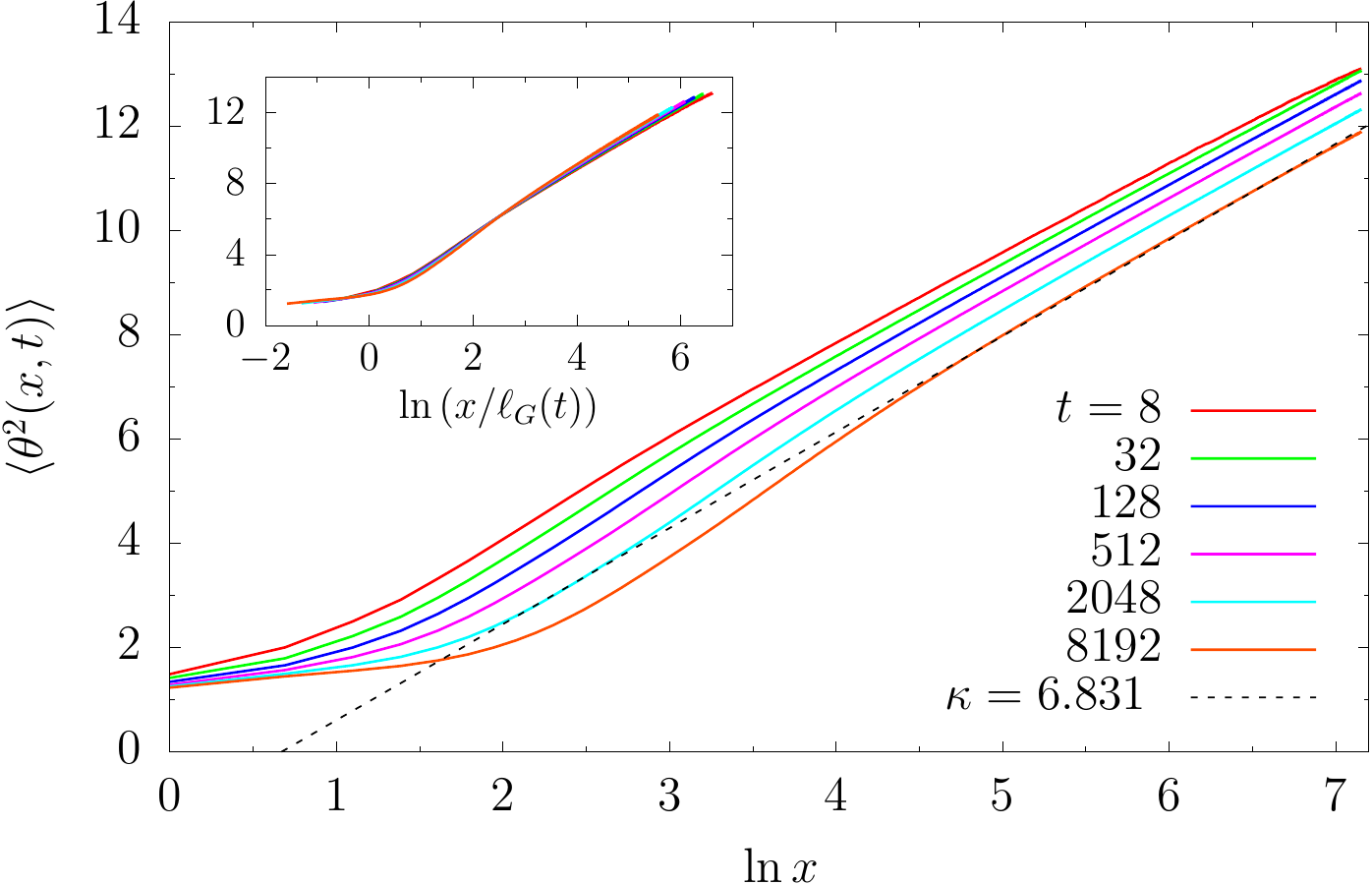}
\end{center}
\caption{\small 
Average squared winding angle $ \langle \theta^2 (x,t) \rangle $ for
the largest cluster interface against
the logarithm of the curvilinear coordinate $x$, at different times given 
in the key after a quench to $T_c/2$.
Ising model with  $L=640$ and equal concentration of up and down spins.
The dotted line is a fit to the data at $t=8192$ that yields $\kappa \simeq 6.831$.
Inset: data vs. $ \ln (x / \ell_G(t))$ with $\ell_G(t)$ the growing 
length measured from the excess energy evaluated at the same times as in the main plot.
}
\label{LKaWA_b2}
\end{figure}

The first observable that we study to determine whether the coarsening dynamics drives the system towards a critical percolating state
is the winding angle, $\theta(x)$, of an interface. It is  defined as the angle between 
the local tangents at two points separated by a curvilinear distance $x$ along the curve
(For a spin configuration on a lattice the local tangent to an interface is a perpendicular vector 
to a broken bond and its direction takes values on a discrete set).
Bidimensional critical interfaces are,  in
the continuum limit, conformally invariant curves described by the stochastic Loewner evolution SLE$_\kappa$.
The average squared $\theta(x)$ can then be exactly computed~\cite{DuSa88,WiWi03}: 
\begin{equation}
\langle \theta^2(x) \rangle = \mbox{cst} +{4 \kappa \over 8+\kappa } \ln{x} 
\; .
\label{ws}
\end{equation}
The parameter $\kappa$ is related
to the interface fractal dimension and determines the universality class.
For critical percolation interfaces, $\kappa = 6$, while for critical Ising ones $\kappa=3$. 

After the quench the averaged variance of the winding angle
approaches the form (\ref{ws})
with $\kappa\simeq 6$
at a time $t_p$, and the later evolution conforms with dynamic scaling 
with the dynamic growing length.
As a proof of this claim, we show  in Fig.~\ref{LKaWA_b2}  
$\langle \theta^2(x,t) \rangle$ for
the largest cluster interface
in a system with $L=640$.
The dotted straight line is 
a fit to the data at $t=8192$ and yields  $\kappa \simeq 6.831$.  
We then used the rescaled curvilinear coordinate $x/\ell_G(t)$ to successfully scale the data at subsequent times, as
shown in the inset.
We also used the data for $\langle \theta^2(x,t) \rangle$ to obtain an alternative estimate of  
$1/z_{\mathrm{eff}}(t)$. We looked for pair-wise collapse of datasets
$ \{ \langle \theta^2(x,t_i) \rangle \}$ at consecutive times with $1/z_{\mathrm{eff}}(t_i)$. The outcome is shown with 
open (green) points in the inset of Fig.~\ref{LKaGL}.

We now turn to the analysis of the time scale $t_p$ and the growing length towards critical percolation
$\ell_p(t)$. In~\cite{BlCuPi14} we used the asymptotic value of the 
overlap $Q(t,t_w) = N^{-1} \sum_{i=1}^N s_i(t) \sigma_i(t_w)$
between two copies of the system, $\{ s_i(t), \sigma_i(t)\}$,
 created at a waiting time $t_w, \, \{s_i(t_w) = \sigma_i(t_w) \}$, that later evolve independently
to measure $t_p(L)$ in an Ising model evolved with Glauber dynamics. The idea was to scale $t_w$ with $L$ and search for the weakest $L$-dependence
in the form $t_w = L^{z_p}$
such that $\lim_{t\gg t_w} Q(t,t_w(L)) >0$ (while it vanishes otherwise). This yields $t_p = t_w(L) =L^{z_p}$.
In the conserved order parameter case, the excess-energy growing length does not 
reach a stable power-law regime and, consequently, it is hard to use a fixed power of $L$ to 
quantify $t_p$. For this reason, we used other observables to estimate $t_p(L)$.

\begin{figure}[h]
\vspace{0.5cm}
\begin{center}
        \includegraphics[scale=0.65]{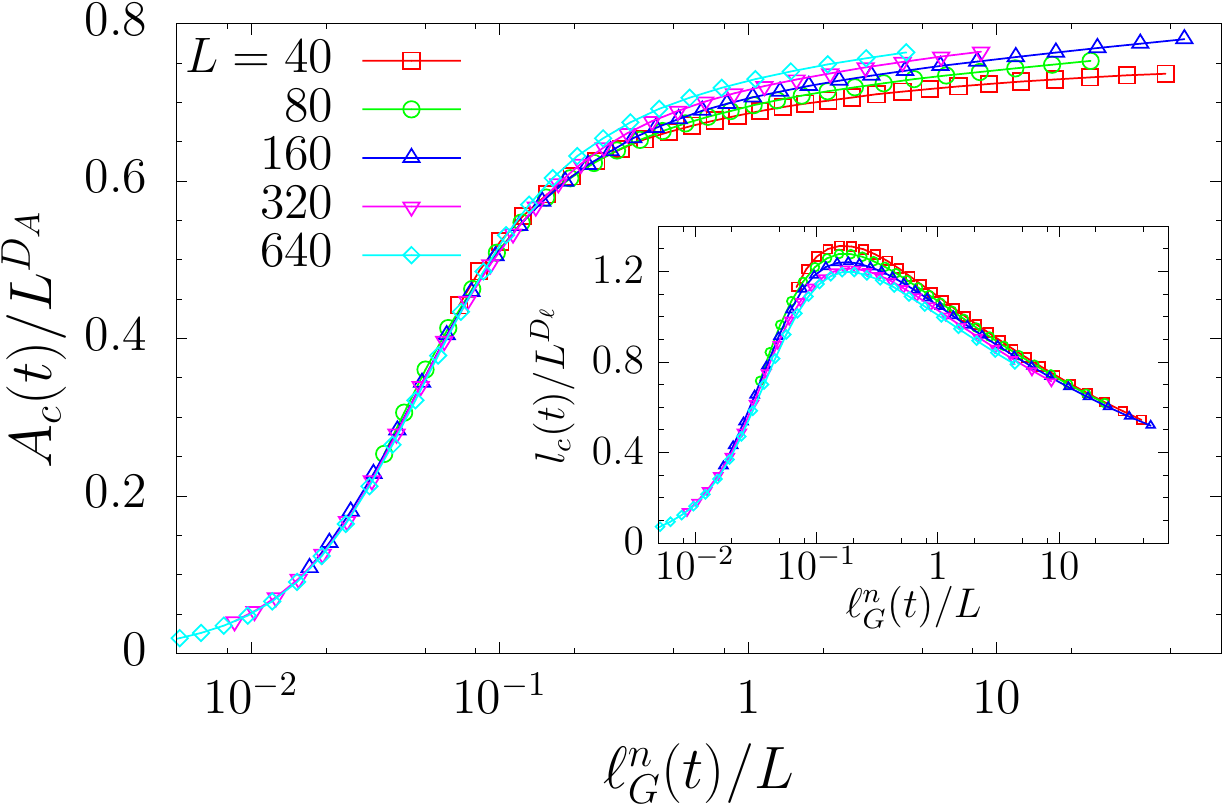}
\end{center}
\caption{\small
Size of the largest cluster $A_c(t)$ divided by $L^{D_A}$ with  $D_A=91/48$
and length of the largest cluster interface, $l_c(t)$, divided by $L^{D_{\ell}}$ with
$D_{\ell}=14/8$,   against the scaling variable $ \ell^n_G(t)/L$. Temperature is $T_c/2$.
The best collapse is obtained with $n\approx 5$, the value used in these plots.
}
\label{LKaLC_b2}
\end{figure}

In  Fig.~\ref{LKaLC_b2} we study the geometric and scaling properties of the 
largest cluster. With the use of 
the area and interface fractal dimensions of critical percolation clusters~\cite{Stauffer94,Christensen02,Saberi15} we find good
scaling of its area and interface length using
$A_c(t)/L^{D_A}$ and $l_c(t)/L^{D_\ell}$ against $\ell^n_{ \rm{G} }(t) / L$ with
$n$ an adjustable parameter. The best collapse
was found using $n=5$, a value that is not far from the guess $n=6$ on the square lattice.
We performed the same analysis on other lattices
and we report here the results 
without showing the scaling plots. On the honeycomb lattice we found
$n_{\rm honey} = 6$ from the scaling of both $A_c$ and $l_c$. 
Instead, on the special triangular case, we could not scale the data for $A_c$ 
while we found acceptable scaling of the data for $l_c$ using $n_{\rm triang} = 6$.

These results
are confirmed by the analysis of
$\pi_{\rm hv}$, $\pi_{\rm h} + \pi_{\rm v}$ and $\pi_{\rm diag}$~\cite{OlKrRe12}
defined as the probability of having a cluster wrapping in both directions of the lattice, 
 in one Cartesian direction only (horizontally or vertically)
and diagonally (as in the fourth sketch of Fig.~\ref{fig:sketch}), respectively. 
These wrapping probabilities have been calculated for critical percolation on a torus in~\cite{Pinson}
and are shown with dotted lines in the upper panel in Fig.~\ref{LKaWrapping_b2}. 
The dynamic data, also shown in this figure against the scaling variable 
$\ell_G^n(t)/L$,
approach these values asymptotically for sufficiently large system sizes.
  Consistently with what we found in the analysis of $A_c$ and $l_c$, the best scaling 
is obtained with $n=5$.  A similar analysis on the triangular and honeycomb lattices 
yield $n_{\rm triang} =n_{\rm honey} = 6$. In all cases other choices of the 
parameter $n$ give scaling plots that are of considerable less quality.

Until now we used an equal concentration of up and down spins. We now search to determine whether the same
critical percolation phenomenon exists for different values of, say, the up spin concentration $p$. In the lower panel in 
Fig.~\ref{LKaWrapping_b2} we display the various wrapping clusters probabilities for $p=0.4, \ 0.42, \ 0.45, \ 0.47, \ 0.5$.
The probability of having a wrapping cluster along one principal direction of the lattice, $\pi_v + \pi_h$ (green curves), 
does not reach the dotted horizontal line for $p = 0.4 < 1-0.5927$ since one of the two species has a percolating cluster initially. For $p=0.42, \ 0.45, \ 0.47$, 
none of the two species percolates initially, and the 
(green) curves increase and approach the critical percolation value but they rapidly detach from it and decrease to zero.
The complementary (red) curves simultaneously approach one.
The curves for $p =  1/2$ are the only ones that approach the non-trivial asymptotes shown with horizontal dotted lines. This result is 
consistent with the observation in~\cite{Takeuchi15}, where the segregating dynamics of a mixture of Bose-Einstein 
condensates was studied.

We stress the fact that $t_p$ is just a characteristic time scale
associated to the approach to the critical percolation state, mostly used in the form $t_p \simeq L^{z_p}$ to scale time as
done quite successfully in~\cite{BlCuPi14}. However it can be useful to have an idea of the order of magnitude
for the system sizes used. One criterium to get a numerical estimate is to look for the time $t$ at which
the wrapping probabilities shown in the top panel of Fig.~\ref{LKaWrapping_b2} reach approximately the asymptotic values
corresponding to critical percolation. This condition corresponds roughly to $\ell^n_G(t_p) / L = 10$ and
yields 
$t_p = 1100, \, 
 2400, \, 
4900, \, 
9400, \, 
17200$ MCs
 for $L= 40, \ 80, \ 160, \ 320, \ 640$, respectively. This relation is approximately linear suggesting $z_p =1$. This value is  in acceptable agreement with 
 the guess $z_p = z_d/n$, if we take into account the fact that the effective $z_d$ is $z_{\rm eff} \simeq 5$, see  Fig.~\ref{LKaGL}, 
 and the measured $n$ is close to $5$.

\begin{figure}[h]
\vspace{0.5cm}
\begin{center}
        \includegraphics[scale=0.6]{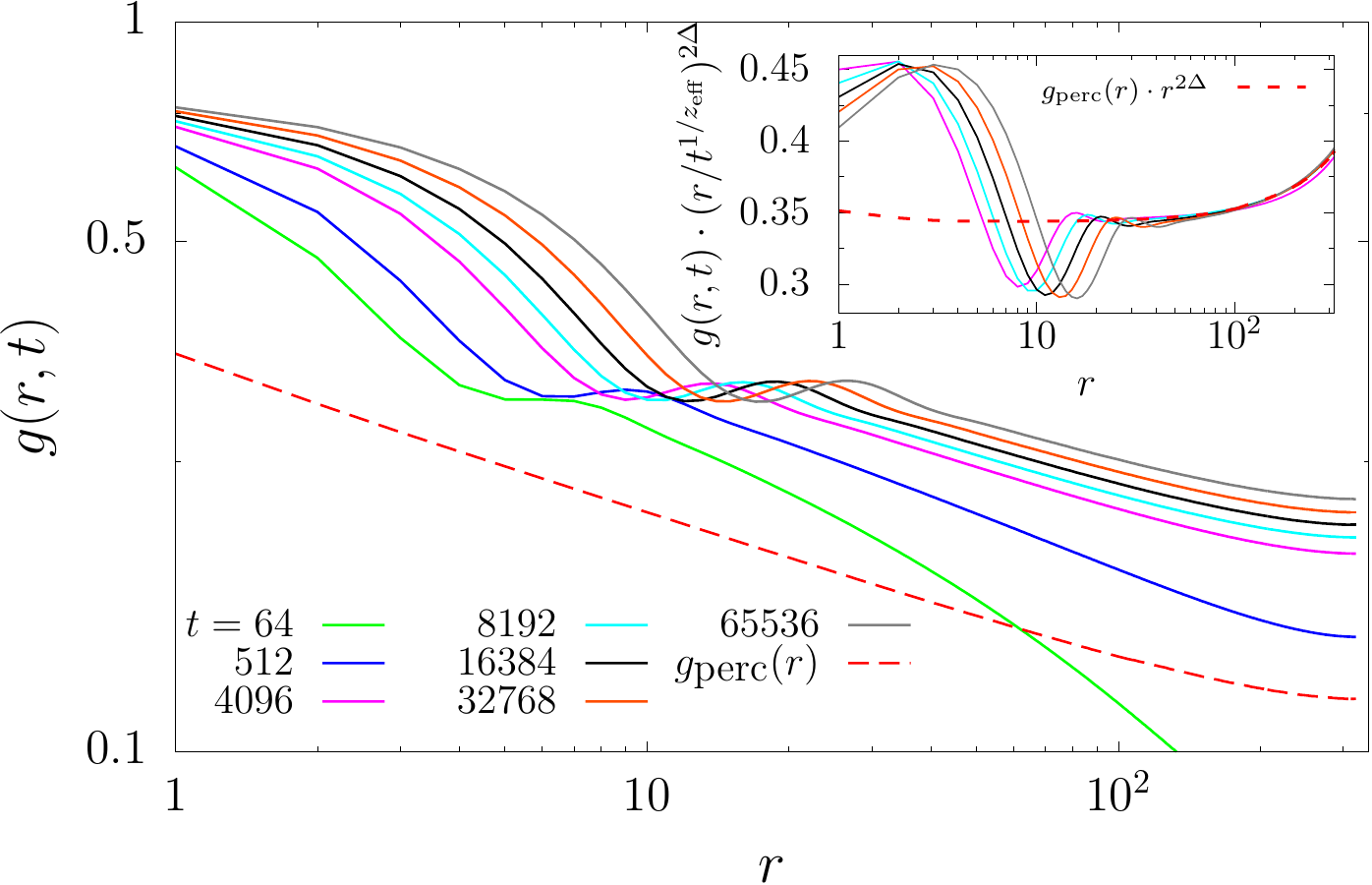}%
\end{center}
\caption{ \small 
Comparison between the time-evolving pair connectedness correlation function and the one for critical site percolation 
shown with a dashed line. $L=640$ and equal concentration of up and down spins. Raw data are shown in the main panel and the
scaling plot in the inset, where  $g(r,t) \cdot (r/t^{1/z_{\rm eff}} )^{2 \Delta}$ is plotted against $r$, with  
$1/z_{\rm eff} = 0.27$ and $\Delta = 2 - D_A$. Temperature is $T_c/2$.
}
\label{LKaPC_b2}
\end{figure}

In  Fig.~\ref{LKaPC_b2}, 
we study the time-dependent pair connectedness correlation function $g(r,t)$ against distance $r$ at different times $t$ 
for a system with $L=640$.
This function measures the probability that two occupied sites (say, with spins 
up) separated at distance $r$ belong to the same cluster~\cite{Stauffer94,Saberi15}. 
We compare the dynamic results to the ones of random site percolation 
at the site occupation probability $p=0.5927$ that is approximately the threshold value $p_c$ (red dashed curve), 
a case for which $g(r) \simeq r^{-2 \Delta}$  where $\Delta = 2 -D_A$ for $a\ll r \ll L$ with $a$ a microscopic length-scale.
Data are for square lattices with the same linear sizes.
In the inset of the same figure we show $ g(r,t) \cdot \, \left[r/t^{1/z_{\rm eff}} \right]^{2\Delta} $ against $r$, 
where we take $1/z_{\rm eff}$ as an adjustable parameter. Using $1/z_{\rm eff} \approx 0.27$
the dynamic data  collapse onto the curve $g_{{\rm perc} } (r) \cdot \, r^{2\Delta}$  of critical percolation
(similarly, for $L=160$ we find $1/z_{\rm eff} \approx 0.28$). The value of $1/z_{\rm eff}$ is
consistent with the effective growth exponent measured from the excess energy, 
see the inset in Fig.~\ref{LKaGL}.
At short length scales, the clusters emerging from the dynamical process are very different from the ones of
critical percolation.
The percolation and dynamic curves are not flat at large $r$ due to the periodic boundary conditions.

The time-dependent number density of geometric domain areas $A$, that we call ${\mathcal N}(A,t, L)$, should scale as 
\begin{equation}
{\mathcal N}(A,t, L) \simeq N(A,t) + N_p(A,t,L)
\label{eq:area-number-density}
\end{equation}
with the first term describing the weight of the finite areas and the second one the 
weight of the areas that span the sample. 
At $t_p$ the last term should become independent of time, and scale with $A/L^{D_A}$. 
A discussion of the scaling properties of the first term 
in terms of  $A/\ell_d^2(t)$, and the crossover of the scaling function from the 
regime in which the scaling variable varies from being much smaller to being much larger
than one was presented in~\cite{SiSaArBrCu09}. In the second regime the first term 
approaches the algebraic finite-size clusters distribution
\begin{equation}
N(A) \, = \, 2 c_d \ A^{-\tau_A}
 \label{eq:na_eq}
\end{equation}
with $\tau_A = 1 + d/D_A$~\cite{Stauffer94,Christensen02,Saberi15}. There is no 
analytic prediction for $c_d$ but a rough estimate suggests
$c_d \simeq 0.029 $~\cite{CaZi03,SiArBrCu07,SiSaArBrCu09}.

\begin{figure}[h]
\vspace{0.5cm}
\begin{center}
        \includegraphics[scale=0.63]{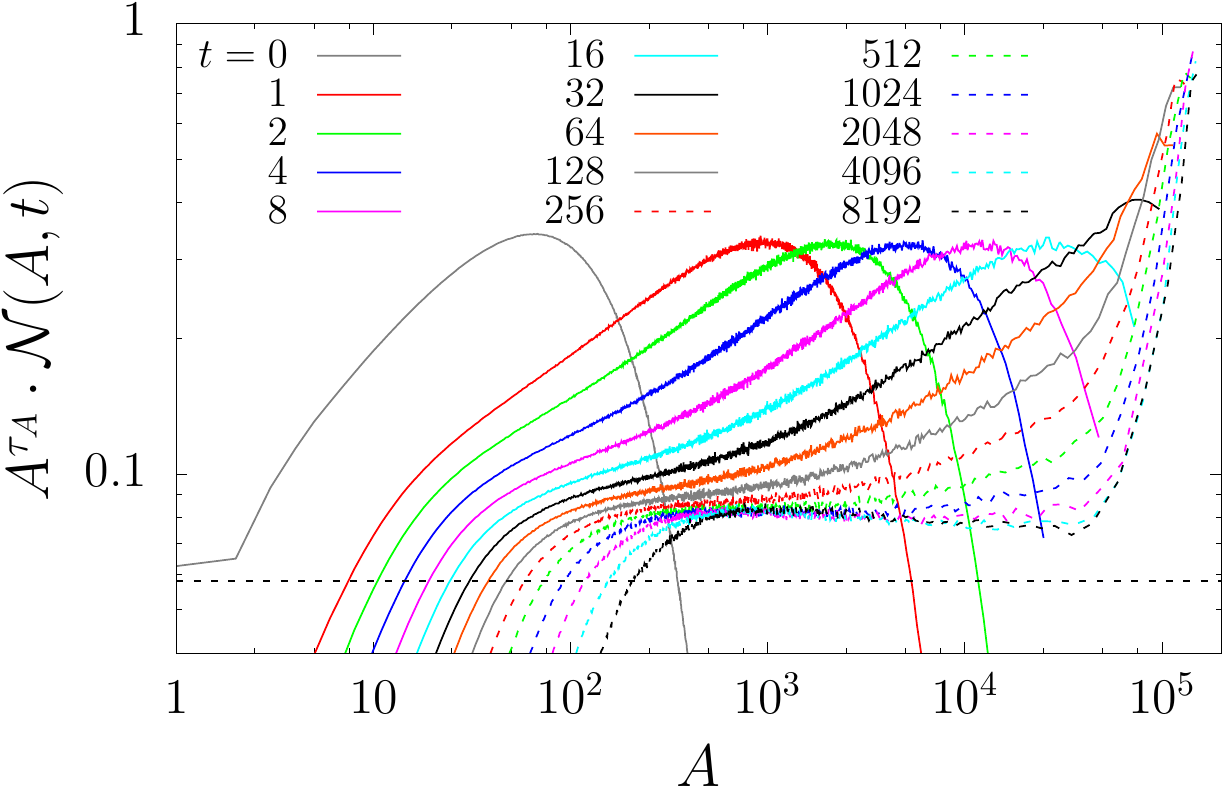}
        \\
        \vspace{0.25cm}
        \hspace{0.2cm}
        \includegraphics[scale=0.6]{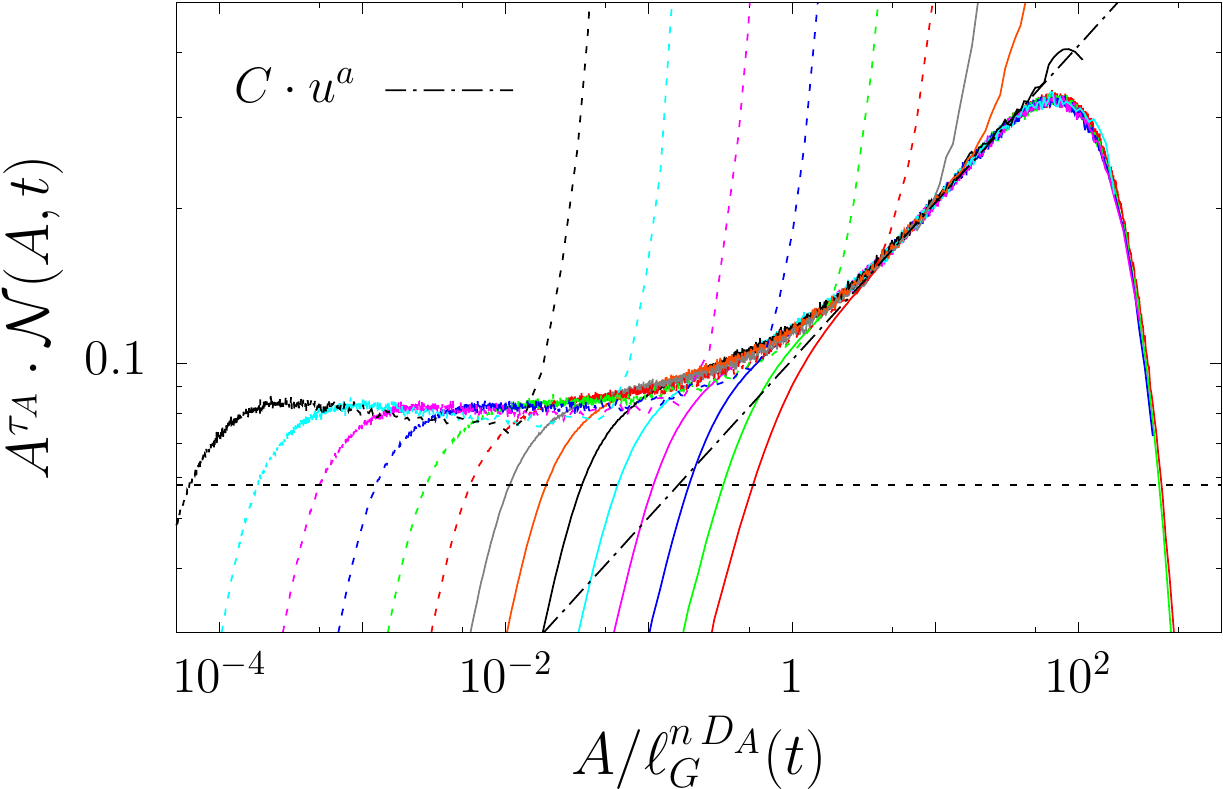}
\end{center}
\caption{\small 
Instantaneous number density of cluster areas  
in a 50:50 mixture with linear size $L=640$ quenched to $T_c/2$.
Upper panel: $A^{\tau_A} \, {\mathcal N}(A,t, L)$, 
with $\tau_A=187/91$ against $A$.
The black horizontal line is at $2 c_d = 0.058 $.
Lower panel: $A^{\tau_A} \, {\mathcal N}(A,t, L)$ against $A/\ell^{n D_A}_G(t)$
with $n=5$ and $\Phi(u) = C \, u^{a}$ with 
$C \simeq 0.101$ and $a \simeq 0.304 $ (dotted line).
}
\label{LKaNA-NWR_b2}
\end{figure}

We first focus on the large area regime of the 
number density of non-wrapping domain areas, that is to say, the 
contribution $N(A,t)$ to the complete number density ${\mathcal N}(A,t,L)$, see Eq.~(\ref{eq:area-number-density}). 
The data for a system with  $L=640$ are presented in Fig.~\ref{LKaNA-NWR_b2},
where $A^{\tau_A} \, {\mathcal N}(A,t,L)$ is plotted against $A$ in the upper panel and against  
$A/\ell^{n D_A}_G(t)$ in the lower panel with $n=5$.
In both cases the critical percolation value 
$\tau_A=187/91$ was used. 

In this representation the flat part is close to $2c_d$~(see~\cite{Tartaglia-long} for more details on the normalisation), 
the vertical looking parts of the curves belong to $N_p$, and the remaining 
bump corresponds to a power-law decay of $N$ with 
an exponent smaller than $\tau_A$. 
This regime is well described by the scaling function 
\begin{equation}
A^{\tau_A} \, {\cal N}(A,t,L) \, \sim \,  \Phi ( A/\ell^{n D_A}_G(t) )
 \label{eq:na_dyn_scaling}
\end{equation}
where $\ell_G^{n D_A}(t)$ has the meaning of a characteristic domain area at time 
$t$, and $\Phi$ is a scaling function such that $\Phi(u) \rightarrow 2c_d$, as
$u \rightarrow 0$, 
and $\Phi(u) = C \,  u^{a}$, with $a>0$.
By fitting this function to the data at early times (up to $t=32$),
in the region $ [1,10^2] $ of the scaling variable $A/\ell^{n D_A}_G(t)$ with $n=5$, we find
$C\simeq 0.101$ and $a \simeq 0.304$. The same scaling behaviour with, interestingly
enough, the same scaling function is found for Glauber dynamics~\cite{Tartaglia-long}.
The collapse of the data for the number density of domain areas is less 
sensitive to the choice of the value of $n$ than the other observables studied. 
For this and the other lattices we found
acceptable scaling for $n$ between $4$ and $6$, approximately.

\begin{figure}[h]
\begin{center}
        \includegraphics[scale=0.6]{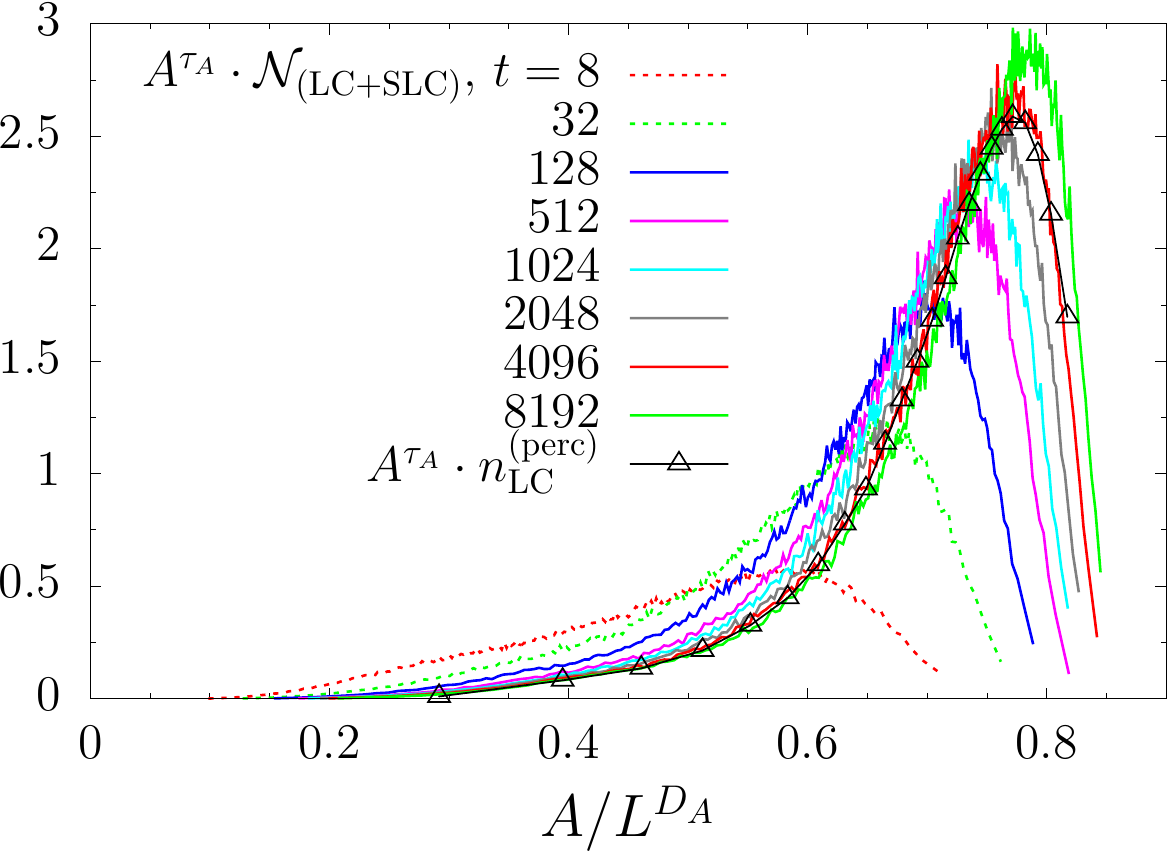}
\end{center}
\caption{\small Size distribution of the largest cluster and second largest cluster, $ \mathcal{N}_{\mathrm{(LC+SLC)}} $,
at different times given in the key in a system with linear size $L=160$, following a sudden quench of a random 
50:50 mixture of up-down spins to $T_c/2$. 
$ A^{\tau_A} \, \mathcal{N}_{\mathrm{(LC+SLC)}}(A,t) $  against
$A/L^{D_A}$, with $\tau_A=187/91$ and $D_A=91/48$.
Also shown is the size distribution, $n^{\rm (perc)}_{\mathrm{LC}}(A)$,  of the largest cluster of random percolation
at $p=0.5927$ that is approximately the threshold value.
}
\label{LKaNA-LC-SLC_b2}
\end{figure}

It is also instructive to study the shape of the probability distribution of the largest clusters. In 
percolation, the distribution of the largest cluster area is a Gumbel distribution
for $p < p_c$~\cite{Bazant00,Sen01,RedigHofstad} while it is approximately Gaussian for $p  > p_c$~\cite{Bazant02}.  
Little is known about its distribution at $p_c$,
except for a remarkable exact result in the mean-field case~\cite{BotetPlos}. 
Numerical studies~\cite{BotetPOS} suggest that 
there is a smooth crossover between the subcritical and the supercritical phase, and that the probability distribution of the 
order parameter can be approximated by a weighted sum of a Gumbel and a Gaussian distribution. The next size in the hierarchy 
is much smaller than the largest one. 

We reckoned in Fig.~\ref{KaSnap} that two very large clusters with size comparable to the system 
size develop dynamically. In Fig.~\ref{LKaNA-LC-SLC_b2} we follow the 
time-evolution of the size distribution of the two largest clusters, $ \mathcal{N}_{\mathrm{(LC+SLC)}} $,
and we compare it to the one of the largest cluster in random site percolation at $p=0.5927$. 
At the late stages of the approach to percolation, $ t \stackrel{<}{\sim} t_p$,   
the distribution of the two largest cluster sizes (after proper rescaling) resembles more and more a Gumbel distribution. 
Beyond $t_p$ the distribution approaches a Gaussian distribution (not shown). 

In conclusion, the ordering dynamics of a phase separating 50:50 mixture starts with  
an approach to critical percolation that lasts for a macroscopic time $t_p$ determined by 
$\ell_p(t_p) \simeq L$, 
when a stable pattern of percolating domains establishes. After this time, the percolating cluster(s) become fatter.
This second regime is characterized by the expected growing length $\ell_d(t) \simeq t^{1/z_d}$ with $z_d=3$.
The growing length  of critical percolation clusters $\ell_p(t)$ is related to the dynamic growing length as determined from the 
excess energy or other observables via Eq.~(\ref{eq:lp-ld}) with the exponent $n$ being numerically close, though not 
identical, to the number of neighbours of a pair of nearest-neighbours on the lattice. 
The approach to percolation and the proper dynamic scaling regime with $z_d=3$ are well-separated, 
similarly to what happens in Glauber dynamics~\cite{BlCuPi14} (where, say on the square lattice, $z_p=1/2$ and $z_d=2$) 
but differently from what was found for the voter model~\cite{TaCuPi15}  (where $z_p \simeq 1.67$ and $z_d=2$).
For unbalanced mixtures in which one species is more present than the other, the dynamics does not reach a long-lasting
critical percolation pattern. 

We have also studied non-local Kawasaki rules in which the pairs of spins updated are not necessarily 
nearest-neighbours finding similar results to the ones discussed here. A detailed comparison with this
and other (Glauber, voter) dynamic rules will be presented in a long publication~\cite{Tartaglia-long}.

\vspace{0.25cm}

\noindent
{\it Acknowledgements.}
We thank F. Corberi and F. Insalata for very useful discussions.
L. F. C. is a member of Institut Universitaire de France.

\bibliographystyle{phaip}
\bibliography{coarsening}

\begin{thebibliography}{10}

\bibitem{Bray94}
A.~J. Bray,
\newblock Adv. Phys. {\bf 43}, 357 (1994).

\bibitem{Bray03}
A.~J. Bray,
\newblock Philos. Trans. Roy. Soc. London {\bf 361}, 781 (2003).

\bibitem{Puri09-article}
S.~Puri,
\newblock Kinetics of phase transitions,
\newblock in {\em Kinetics of Phase transitions}, edited by S.~Puri and
  V.~Wadhawan, Taylor and Francis, 2009.

\bibitem{GonnellaYeomans09}
G.~Gonnella and J.~Yeomans,
\newblock Using the lattice {B}oltzmann algorithm to explore phase ordering in
  fluids,
\newblock in {\em Kinetics of Phase Transitions}, edited by S.~Puri and
  V.~Wadhawan, Taylor and Francis, 2009.

\bibitem{Tanaka12}
H.~Tanaka,
\newblock Phase separation in soft matter: concept of dynamic asymmetry,
\newblock in {\em Soft Interfaces}, edited by D.~Quer\'e, L.~Bocquet,
  T.~Witten, and L.~F. Cugliandolo, Les Houches Summer School XCVIII, Oxford
  University Press, 2015.

\bibitem{CorberiPoliti}
F.~Corberi and {P. Politi~eds.},
\newblock {\em Coarsening dynamics},
\newblock Comptes Rendus de Physique {\bf 16}, 255 (2015).

\bibitem{De-etal14}
S.~De et~al.,
\newblock Phys. Rev. A {\bf 89}, 033631 (2014).

\bibitem{Tojo-etal}
S.~Tojo et~al.,
\newblock Phys. Rev. A {\bf 82}, 033609 (2010).

\bibitem{Hoffman-etal14}
J.~Hofmann, S.~S. Natu, and S.~{Das~Sarma},
\newblock Phys. Rev. Lett. {\bf 113}, 095702 (2014).

\bibitem{KudoKawaguchi}
K.~Kudo and Y.~Kawaguchi,
\newblock Phys. Rev. A {\bf 88}, 013630 (2013).

\bibitem{Takeuchi15}
H.~Takeuchi, Y.~Mizuno, and K.~Dehara,
\newblock Phys. Rev. A {\bf 92}, 043608 (2015).

\bibitem{Takeuchi16}
H.~Takeuchi,
\newblock J. Low Temp. Phys. {\bf 183}, 169 (2016).

\bibitem{LifshitzSlyozov59}
I.~M. Lifshitz and V.~V. Slyozov,
\newblock Zh. Eksp. Teor. Fiz. {\bf 35}, 479 (1959).

\bibitem{Wagner61}
C.~Wagner,
\newblock Z. Elektrochem. {\bf 65}, 581 (1961).

\bibitem{Hu86}
D.~A. Huse,
\newblock Phys. Rev. B {\bf 34}, 7845 (1986).

\bibitem{Amar88}
J.~G. Amar, F.~E. Sullivan, and R.~D. Mountain,
\newblock Phys. Rev. B {\bf 37}, 196 (1988).

\bibitem{Rogers88}
T.~M. Rogers, K.~R. Elder, and R.~C. Desai,
\newblock Phys. Rev. B {\bf 37}, 9638 (1988).

\bibitem{Reith12}
D.~Reith, K.~Bucior, L.~Yelash, P.~Virnau, and K.~Binder,
\newblock J. Phys.: Condens. Matter {\bf 24}, 115102 (2012).

\bibitem{SiSaArBrCu09}
A.~Sicilia, Y.~Sarrazin, J.~J. Arenzon, A.~J. Bray, and L.~F. Cugliandolo,
\newblock Phys. Rev. E {\bf 80}, 031121 (2009).

\bibitem{ArBrCuSi07}
J.~J. Arenzon, A.~J. Bray, L.~F. Cugliandolo, and A.~Sicilia,
\newblock Phys. Rev. Lett. {\bf 98}, 145701 (2007).

\bibitem{SiArBrCu07}
A.~Sicilia, J.~J. Arenzon, A.~J. Bray, and L.~F. Cugliandolo,
\newblock Phys. Rev. E {\bf 76}, 061116 (2007).

\bibitem{BlCoCuPi14}
T.~Blanchard, F.~Corberi, L.~F. Cugliandolo, and M.~Picco,
\newblock EPL {\bf 106}, 66001 (2014).

\bibitem{Kawasaki66a}
K.~Kawasaki,
\newblock Phys. Rev. {\bf 145}, 224 (1966).

\bibitem{Kawasaki66b}
K.~Kawasaki,
\newblock Phys. Rev. {\bf 148}, 375 (1966).

\bibitem{Bortz-etal74}
A.~B. Bortz, M.~H. Kalos, J.~L. Lebowitz, and M.~A. Zendejas,
\newblock Phys. Rev. B {\bf 10}, 535 (1974).

\bibitem{Barkema}
G.~T. Barkema and M.~E.~J. Newman,
\newblock {\em Monte Carlo methods in statistical physics},
\newblock Oxford University Press, Oxford, 1999.

\bibitem{Tartaglia-long}
T.~Blanchard, L.~F. Cugliandolo, M.~Picco, and A.~Tartaglia,
\newblock {\it Critical percolation in bidimensional kinetic spin models}, in
  preparation, 2016.

\bibitem{DuSa88}
B.~Duplantier and H.~Saleur,
\newblock Phys. Rev. Lett. {\bf 60}, 2343 (1988).

\bibitem{WiWi03}
B.~Wieland and D.~B. Wilson,
\newblock Phys. Rev. E {\bf 68}, 056101 (2003).

\bibitem{BlCuPi14}
T.~Blanchard, L.~F. Cugliandolo, and M.~Picco,
\newblock J. Stat. Mech. , P12021 (2014).

\bibitem{Stauffer94}
D.~Stauffer and A.~Aharony,
\newblock {\em Introduction To Percolation Theory},
\newblock Taylor and Francis, London, 1994.

\bibitem{Christensen02}
K.~Christensen,
\newblock {\it Percolation Theory},
\newblock available at
  \url{http://www.mit.edu/~levitov/8.334/notes/percol_notes.pdf}, 2002.

\bibitem{Saberi15}
A.~A. Saberi,
\newblock Phys. Rep. {\bf 578}, 1 (2015).

\bibitem{OlKrRe12}
J.~Olejarz, P.~L. Krapivsky, and S.~Redner,
\newblock Phys. Rev. Lett. {\bf 109}, 195702 (2012).

\bibitem{Pinson}
H.~Pinson,
\newblock J. Stat. Phys. {\bf 75}, 1167 (1994).

\bibitem{CaZi03}
J.~Cardy and R.~M. Ziff,
\newblock J. Stat. Phys. {\bf 110}, 1 (2003).

\bibitem{Bazant00}
M.~Z. Bazant,
\newblock Phys. Rev. E {\bf 62}, 1660 (2000).

\bibitem{Sen01}
P.~Sen,
\newblock J. Phys. A {\bf 34}, 8477 (2001).

\bibitem{RedigHofstad}
R.~van Der~Hofstad and F.~Redig,
\newblock J. Stat. Phys. {\bf 122}, 671  (2006).

\bibitem{Bazant02}
M.~Z. Bazant,
\newblock Physica A {\bf 316}, 29  (2002).

\bibitem{BotetPlos}
R.~Botet and M.~Ploszajczak,
\newblock Phys. Rev. Lett. {\bf 95}, 185702 (2005).

\bibitem{BotetPOS}
R.~Botet,
\newblock Proceedings of Science (WPCF2011) {\bf 007} (2012).

\bibitem{TaCuPi15}
A.~Tartaglia, L.~F. Cugliandolo, and M.~Picco,
\newblock Phys. Rev. E {\bf 92}, 042109 (2015).

\end{thebibliography}

\end{document}